# Cyclic game dynamics driven by iterated reasoning

Seth Frey[a,1], Robert L. Goldstone[a]

**Affiliations:**
[a] *Cognitive Science, Indiana University,*
*1101 E. 10th St., Bloomington, IN, 47405, USA.*

**Corresponding author:**
[1] Seth Frey
Cognitive Science, Indiana University,
1101 E. 10th St., Bloomington, IN, 47405, USA.
Phone: (812) 567-3674
Fax: (812) 855-4691
Email: sethfrey@indiana.edu

**Abstract**

Recent theories from complexity science argue that complex dynamics are ubiquitous in social and economic systems. These claims emerge from the analysis of individually simple agents whose collective behavior is surprisingly complicated. However, economists have argued that iterated reasoning—what you think I think you think—will suppress complex dynamics by stabilizing or accelerating convergence to Nash equilibrium. We report stable and efficient periodic behavior in human groups playing the Mod Game, a multi-player game similar to Rock-Paper-Scissors. The game rewards subjects for thinking exactly one step ahead of others in their group. Groups that play this game exhibit cycles that are inconsistent with any fixed-point solution concept. These cycles are driven by a "hopping" behavior that is consistent with other accounts of iterated reasoning: agents are constrained to about two steps of iterated reasoning and learn an additional one-half step with each session. If higher-order reasoning can be complicit in complex emergent dynamics, then cyclic and chaotic patterns may be endogenous features of real-world social and economic systems.

**Introduction**

When seen at the level of the entire group, the reasoning of many individuals can lead to unexpected collective outcomes, like wise crowds, market equilibrium, or tragedies of the commons. In these cases, people with limited reasoning can converge upon the behavior of rational agents. However, limited reasoning can also reinforce dynamics that do not converge upon a fixed-point. We show that bounded iterated reasoning through the reasoning of others can support a stable and profitable collective behavior consistent with the limit cycle regimes of many standard models of game learning.

A limit cycle is a set of points within a closed trajectory, and it is among the simplest non-fixed-point attractors. Game theorists have been demonstrating the theoretical existence of limit cycle attractors since the 1960s (Shapley, 1964) and cyclic dynamics have been identified in every classic learning model (Fudenberg & Kreps, 1993; Hommes & Ochea, 2011; Jordan, 1993; Sato, Akiyama, & Crutchfield, 2005). In some models, cyclic regimes emerge when payoff (or sensitivity to it) is low (Sato & Crutchfield, 2003). Theorists, particularly those interested in the replicator dynamic, have also discovered more complex attractors in belief space, like chaos in simple and complex games (Galla & Farmer, 2011; Sato, Akiyama, & Farmer, 2002). Kleinberg et al. (2011) remind us that cyclic learning dynamics may be more efficient than those that converge to a fixed point.

Should we expect similar complexity in actual human behavior? Humans are capable of "higher" types of reasoning that are absent from most theoretical models, and that have not been empirically implicated in complex dynamics. In work to demonstrate the stabilizing role of iterated reasoning, Selten proved that for a large class of mixed-strategy games, and sufficiently slow learning, adding iterated reasoning to a simple replicator dynamic guarantees the local stability of all Nash equilibria (1991). Behavioral experiments have supported the thrust of this claim (Bloomfield, 1994; Tang, 2001) and, in work with a similar motivation, Camerer et al. showed that iterated reasoning can help standard adaptive learning models converge to Nash equilibrium at the rates observed in humans (2002a).

Cyclic game dynamics have been observed in organisms that are not capable of higher-order reasoning. Animal behavior researchers have described the role of periodic dynamics in resolving coordination conflicts in the producer-scrounger problem (Rands, Cowlishaw, Pettifor, Rowcliffe, & Johnstone, 2003; Sumpter, 2010, p. 149). Rock-Paper-Scissors relations, and cycles within them, have been identified among side-blotched lizards and *in vitro* and *in vivo* populations of E. Coli, and they have been implicated in the maintenance of species diversity (Alonzo & Sinervo, 2001; Kerr, Riley, Feldman, & Bohannan, 2002; Kirkup & Riley, 2004; Reichenbach, Mobilia, & Frey, 2007).

When experimentalists entertain dynamic models of human behavior, they tend to treat non-Nash behavior as part of the process of eventually converging to Nash (Camerer, 2003, p. 270). However, experiments in games that prescribe random (mixed-strategy) play document sustained distance from predicted equilibria and/or failure to converge to a fixed point (Bloomfield, 1994; Bottazzi & Devetag, 2007; Brown & Rosenthal, 1990; Crawford & Iriberri, 2007a; Duffy & Hopkins, 2005; Hoffman, Suetens, Nowak, & Gneezy, 2012; Rapoport & Budescu, 1992; Rubinstein & Tversky, 1993). These studies account for their results by citing cognitive limits, poor motivation, or by resorting to alternative, sometimes unspecified, solution concepts. However, there is also positive behavioral evidence for specific higher-dimensional attractors. One example is the Edgeworth cycle in duopolistic markets (Cason, Friedman, & Wagener, 2005; Eckert & West, 2004; Edgeworth, 1925; Noel, 2007; Zhang & Feng, 2005), though its mechanism does not invoke learning or implicate higher-order reasoning. Another example is the hog cycle that motivated rational expectations theory (Muth, 1961). Recently, experimentalists have been observing cyclic choice dynamics in the lab as well (Bednar, Chen, Liu, & Page, 2012; Cason, Friedman, & Hopkins, 2012).

We introduce the Mod Game, an *n*-player generalization of Rock-Paper-Scissors. Its name evokes a couturier's designs to anticipate the recurrence of previously outmoded fads within a peer community. Behavior in the game is inconsistent with any fixed-point attractor concept, and consistent with the long history of predictions of cyclic attractors in game learning. This result comes with evidence for iterated reasoning through the reasoning of others, and with the emergence of self-organized clustering.

**Methods**

*The Mod Game*

In the Mod Game, *n* participants choose an integer in the range {1, …, *m*} for *n* and *m* both greater than one. Every participant earns a point for each choice by another that they exceeded by exactly one; e.g., Choice 3 dominates (or "beats") Choice 2 (and only Choice 2), and Choice 2 beats Choice 1. The exception to this scoring rule is that Choice 1 beats Choice *m*. This exception gives the game the intransitive dominance structure of Rock-Paper-Scissors, in which there is no single action that cannot be dominated by some other action. All players in a group play against all others simultaneously each round, so a player beating two others receives two points, and two players each earn one point if they both chose the same choice and beat a third player. A player whose choice is not exactly one more than another's scores zero points. The game is not zero-sum and players do not lose any points for making choices that benefit other group members.

In our implementation, the maximum integer choice *m* equaled 24. At *m* = 3, the game is a non-zero-sum version of Rock-Paper-Scissors. Experimentalists have observed cyclic dynamics in intransitive games with *m* equal to two and three (Bednar et al., 2012; Cason et al., 2012). However, larger values of *m* permit greater discrimination between potential reasoning processes behind behavior. For example, with only three strategies it is very difficult to determine whether Rock, as a response to Scissors, is the result of zero, three, or even six levels of iterating reasoning. By increasing the number of cyclically arranged choices to 24, we can determine the order of iterated reasoning with less ambiguity.

*Experimental design*

After all decisions were submitted, all of the round's choices and earnings were revealed to all players, and the game was repeated for 200 rounds. We also tested a symmetric condition (*decrement*) in which the scoring rule was reversed and players were rewarded for choices exactly one less than those of other participants, with the exception of Choice 24, which rewarded one point for each group member that selected Choice 1. This second condition helped distinguish the effects of the scoring rule from other possible incidental effects of the experimental environment.

*Procedure*

Over 22 sessions at Indiana University, 123 psychology undergraduates played in groups of 2–10. The scoring rule does not demand a specific group size, and our design only controlled for group size statistically. Figures S2-S9 summarize the complete data from the experiment. Table 1 lists the group sizes for each session. Participants were instructed to earn as many points as possible. In addition to course credit for appearing at the experiment, they were given a cash bonus based on the number of points they earned over all rounds. Specifically, one of every ten rounds was randomly selected as a "pay round" in which participants were rewarded 10¢ for each point. In all rounds, a participant has six seconds to make a non-null decision. Six seconds was ample time for most participants; only 1.5% of decisions were null. The mean session lasted 24 minutes.

Subjects sat at curtained terminals, and interacted with a graphical Java-based interface using the HubNet plugin for NetLogo (Wilensky, n.d.; Wilensky & Stroup, n.d.). After the experiment administrator read the instructions publicly, subjects were given time to read the text of the instructions individually.

> You are playing a game with other people. Your goal is to earn as many points as possible. Everyone in your group will choose from a circle of numbered squares 200 times. Your goal is to choose a square that is one more [less] than other people's squares. The squares wrap around so that the lowest [highest] choice counts as just above the highest [lowest] (like an ace sometimes counts as higher than a king, but still below a two). You get one point for every person who you are above [below] by only one square.
>
> As a bonus, you will be paid for earning as many points as you can. We will pick twenty random rounds and pay you 10 cents per point.

The experiment began after all participants finished reviewing the instructions. Subjects' 24 strategies were arrayed visually in a circle (Figure 1). To distinguish the potential visual salience of specific choices (e.g. the highest and lowest numbers 1 and 24) from that of specific screen locations (e.g. the top-, bottom-, and right-most choices), each group was presented with a circle whose choices had been rotated by a different random amount at the initialization of the experiment. Averaging over all rounds and sessions, participants showed mild preferences for choices 1, 7, 24, and no particular preference for any visual location on the circle. Figure 1 shows the graphical interface of the game.

Though participants were instructed to earn as many points as possible, some exhibited behavior that could not have assisted them towards this end. In particular, some participants repeated their previous round's choice for large parts of the experiment. Of an original 167 participants, 8 had "streaks" of the same choice for 25 or more rounds in row (1/8 of the total experimental session). In group experiments, individuals influence their group's behavior, so we cautiously threw away all 8 experiments in which these 8 subjects had participated. The resulting subject pool had 123 participants. The discussion will explore questions of motivation and robustness but, in summary, the results we report are robust to an analysis that includes all 167 participants, and the complete discarded data are available for inspection in Figures S8 and S9.

*Ethics Statement*

This manuscript reports experimental data from human subjects. Written informed consent was obtained after the nature and possible consequences of the studies were explained. The research contained in this submission was approved by the Indiana University Institutional Review Board.

*Measures*

In games with mixed-strategy Nash equilibria, there is prior experimental evidence for two related-but-distinct outcomes: a failure to converge to some fixed-point solution concept (like Nash equilibrium) and a failure to converge to any fixed-point solution concept. These can be established in a Mod Game with an assortment of complimentary dependent measures. Other methods, like frequency analysis, can then be used towards supporting alternatives to fixed-point convergence.

We used participant time series—vectors of 200 integers valued 1 through 24—to measure *entropy*, *efficiency*, *distance*, and two measures of sequential dependence, *rate* and *acceleration*. *Entropy* is the information entropy of each individual's time series (Shannon, 1948). Information entropy is a measure of disorder in distributions, such that samples from uniform distributions offer the least information per observation. This measure can be used to compare the disorder in observed behavior to that of a random benchmark. For each participant *i*, information entropy, $H(X_i)$, was calculated from the empirical probability distribution function of random variable $X_i$, which can take the 24 possible values of $x_j$, with $H(X_i) = -\sum_{j=1}^{24} p(x_j = X_i) \log p(x_j = X_i)$.

*Efficiency* is the percentage of points scored in a round, out of the maximum possible for that group size. Efficient behavior in the Mod Game is profitable behavior, and is an implicit measure of the effectiveness of groups to coordinate for greater gains. Efficiency $E$ was measured for each round $t$, as $E(t) = \dfrac{\pi(t)}{(\lfloor n/2 \rfloor \lceil n/2 \rceil)}$, where $\pi(t)$ is the sum of points earned at $t$, and $n$ is the group size. The denominator gives the maximum possible number of points within a round of the game; Efficiency is constrained to the [0, 1] interval. Maximum efficiency can be achieved if half of the members of a group (or about half, for odd group sizes) select one choice, and the other half select a choice exactly one above or below.

We introduced *distance* to measure the clustering of choices within rounds. Clustering is a type of coordination that has been observed in similar environments (S. Frey & Goldstone, 2011). Taking the distance between two participants as the shortest path between them on the circle of choices, the value of distance in a round is the mean of the distances between all pairs of choices in that round. Low values of distance imply more clustering of choices within a round. Distance $D_i(t)$ was measured for each subject at each round *t*. Subject *i*'s choice in a round is denoted by $s_i(t)$, and the choices of the other group members are $S_{-i}(t)$. $D_i(t) = \dfrac{1}{n} \sum_{j \in S_{-i}(t)} \min(|b-a|, |(a+24)-b|)$ where *a* and *b* are $\min(s_i(t), j)$ and $\max(s_i(t), j)$. This function identifies the shortest paths between choices 5 & 7 and 1 & 23 as having distance 2, rather than 22. A round's distance $D(t)$ was a mean of individual distances, $D(t) = \dfrac{1}{n} \sum_{i=1}^{n} D_i(t)$.

The last two measures gave insight into sequential dependence—how a choice in one round predicts choices in future rounds. While series of random choices should be statistically independent, past experiments in games with intransitive dominance have documented significant sequential dependencies, usually attributed to cognitive or motivational limits (Bottazzi & Devetag, 2007; Duffy & Hopkins, 2005; Rapoport & Budescu, 1992).

We tested for sequential dependence with analyses of the distributions of first and second differences of participant time series, what we define as *rate* and *acceleration*. We calculated rate as the time series of 199 differences between consecutive raw choices, modulo 24. The modulus was taken to define rate on the interval {0, ..., 23}. The second difference is the sequence of 198 differences between consecutive first differences, also converted to the interval {0, ..., 23}. Under random behavior, these constructs should be uniformly distributed, like the raw choices from which they are calculated. These tests of dependence motivated further tests for periodicity in the observed behavior.

*Predictions*

**Hypothesis 1:** Behavior in the Mod Game will be consistent with uniformly random behavior.

The Mod Game is intransitive in that there is no single action that cannot be dominated by another; the game has no pure-strategy Nash equilibrium. For group sizes that are not evenly divisible by twenty-four, and for all of the group sizes we tested, the unique Nash equilibrium is to randomly choose from the 24 choices uniformly. This mixed-strategy equilibrium may seem to be a very naïve null model of actual human behavior — Botazzi and Devetag observe that random play is only rational when others are expected to play randomly (2007). However, more recent and psychologically plausible solution concepts *also* predict uniformly random behavior in the Mod Game (e.g. the Cognitive Hierarchy of Camerer, Ho, & Chong, 2002b, and quantal response equilibrium, McKelvey & Palfrey, 1995).

Hypothesis 1 can be rejected using any of the first three measures. Compute benchmark values of entropy, efficiency, and clustering for simulated agents with uniformly random behavior. Though baseline entropy is simple to compute by hand, the other two measures have different baseline values for different group sizes, and simulation was more convenient. If observed values of these measures are significantly different from random benchmark values, Hypothesis 1 can be rejected.

Rejecting Hypothesis 1 would not be particularly provocative. Deviations from uniformly random behavior, which are typical at the individual level anyway, are as likely to result from individual cognitive limits as from convergence upon a higher dimensional attractor.

**Hypothesis 2:** Behavior in the Mod Game will be consistent with some fixed-point of a learning dynamic.

This hypothesis can be rejected by looking at sequential dependence. Even if participants do not converge upon uniformly random play, they may have settled upon some other, possibly less principled mixed-strategy. Significant sequential dependence (or a meaningful *rate* in the terms above) is incompatible with mixed-strategy play; if the distribution of observed rates is significantly different from a uniform distribution over

{1, …, 24}, than we can reject Hypothesis 2, that behavior in the Mod Game is consistent with convergence to a fixed-point.   However, as with Hypothesis 1, rejecting this second Hypothesis is not particularly provocative.   Non-convergent dynamics have been isolated in iterated games, particularly in games with mixed-strategy equilibria.

**Hypothesis 3:** Behavior in the Mod Game will be consistent with the convergence of beliefs towards a periodic attractor.

This hypothesis is motivated by observations, in every major class of learning model, of cyclic attractors in games with mixed-strategy equilibria. Supporting Hypotheses 1 or 2 precludes support for Hypothesis 3.

Many high dimensional attractors can exhibit periodicity. While the most common is the limit cycle, this Hypothesis does not specify an attractor, merely that it will have periodic dynamics.   Periodicity can be established with Fourier analysis, though it takes statistical methods peculiar to frequency space to distinguish a specific frequency component, or an entire spectrum, from white noise.

**Results**

**Result 1: Behavior was inconsistent with uniformly random mixed-strategies.** The entropy expected from random play was above the 99% confidence interval for observed entropy (Figure 2). Both efficiency and distance measures suggest that participant's choices were statistically dependent upon each other. Mean efficiency was significantly higher than that expected from random behavior, and participants' choices clustered significantly by round.

**Result 2: Behavior was inconsistent with convergence to any fixed-point.** A participant's behavior in a given round was also dependent on their behavior in the previous round. Figure 3 shows the distribution of observed and randomized choices, rates, and accelerations, over both conditions. Participants tended to select a choice 4-8 choices "ahead" of their previous choice (modulo 24, and "behind" for subjects in the *decrement* condition). Sequential changes to this rate were small; 53.2% of accelerations—over 24,043 individual decisions—either maintained the previous round's rate or stayed within two choices of it.

**Result 3: Behavior was consistent with convergence to a periodic attractor.** If rate is a meaningful construct in this game, whose strategies are arranged in a circle, then stable rate implies stable periodicity. If participants cycle stably around the strategy set (the circle of choices), any periodicity will show in a Fourier decomposition of their choice sequences. A frequency spectrum may exhibit a larger component at the frequency predicted by the mean rate of rotation.

Since the time series of participants in a group are dependent on each other, data were resampled prior to the frequency analysis. We bootstrapped an independent distribution of observations by randomly selecting one participant's time series from each of the (statistically independent) groups, and we repeated this sampling procedure 1000 times. Each resulting time series was transformed to the frequency domain with the FFT. Before this operation, missing choices (from the 1.5% of rounds in which an individual made no entry, leaving 24,034 of 24,400 data points) were cautiously replaced with uniform noise from the integer interval {1, …, 24}. Reported spectra and confidence intervals were estimated from this large bootstrapped sample of spectra. The white noise registers artificially low amplitude at frequency zero because of how the data were normalized for transfer to the frequency domain.

We combined data from the *increment* and *decrement* conditions by artificially "flipping" all data in the *decrement* condition to exhibit positive rotation, as in $f(x) = -x \mod 24$. Because phase information is discarded in the analysis of frequency spectra, this manipulation should not compromise the analysis.

Data were also transformed prior to the frequency analysis. Because of the "jump" between Choices 1 and 24, any cycles around the raw choices describe a sawtooth curve. Sawtooth curves exhibits well-documented artifacts in frequency spectra, such that a sawtooth with fixed frequency will register many components after decomposition by the Fourier method. To control these artifacts prior to frequency analysis, each time series was transformed to represent the *shortest* distance from an arbitrary fixed point on the circle of strategies (e.g. Choice 1); for Choice *x* scaled to the interval [-1,1], $f(x) = |2x| - 1$. This alternative representation varies without the large periodic discontinuities that characterize sawtooth curves, and the component for the basic frequency of a transformed sawtooth curve be attended by fewer artifactual components.

We then conducted a distributional test in the frequency domain as a preliminary test for periodicity (Figure 4). The Box-Ljung Q test examines statistical features of an autocorrelation to test the null hypothesis of sequential independence in a time series. The test statistic is $\chi^2$ distributed, with degrees of freedom equal to the number of lags, such that with 10 lags the null value of the statistic is equal to 10. A bootstrapped distribution of observed values of the statistic had a mean of 36.6, over 99% CI [36.5, 36.7]. Under this test, we rejected the hypothesis that observed power spectra were generated by random series ($\chi^2_{10}$=33.6, p<0.001).

We complemented the distributional test with a point test for stable periodic behavior at a predicted frequency. This prediction was based on the mean rate of rotation, estimated as the mean of a *von Mises* distribution fit to the histogram in panel 2 of Figure 3. The von Mises distribution is a circular analogue of the normal distribution and it is apt because a) a rate of 0 is equidistant from rates 1 and 23 and b) if an observed rate of *x* corresponds to intended motion at all, it may reflect an intention to advance by *x* plus any integer multiple of twenty-four (including intended motion "backwards").

As fit to a von Mises distribution, the maximum-likelihood mean rate was 4.7 choices per round, corresponding to a predicted frequency of 0.2 rotations per round. A bootstrapped empirical distribution of the amplitude of the 0.2 frequency component placed it above the amplitude expected from random behavior (mean 1.06, 99% CI [1.04, 1.08], above the amplitude of noise at 0.82).

Video of a typical session gives a subjective associate to the statistical support for periodicity (Figure S1). Video also shows that rotation and clustering seem to emerge together, two facets of the same phenomenon.

**Result 4: Rate of rotation increased with time.**

We used linear mixed effects to test potential modulators of participant rate. Our model of rate

$$Rate_{i,subj,group} = \beta_0 + \beta_{round} + \beta_{groupsize} + \beta_{condition} + (u_{subj} + u_{group} + e_{i,subj,group})$$

controlled for both individual- and group-level differences, modeled as random effects $u_{subj}$ and $u_{group}$. $\beta_{round}$ and $\beta_{groupsize}$ fit for the effects of time (with values $\{1, \ldots, 200\}$) and group size. $\beta_{condition}$ fit any difference between the increment and decrement conditions. We compared this model with the three reduced models

$$Rate_{i,subj,group} = \beta_0 + \beta_{round} + \beta_{groupsize} + (u_{subj} + u_{group} + e_{i,subj,group})$$
$$Rate_{i,subj,group} = \beta_0 + \beta_{round} + \beta_{condition} + (u_{subj} + u_{group} + e_{i,subj,group})$$
$$Rate_{i,subj,group} = \beta_0 + \beta_{groupsize} + \beta_{condition} + (u_{subj} + u_{group} + e_{i,subj,group})$$

These tests supported the indifference of rate to group size and condition, and rejected the null hypothesis that rate is indifferent to round (Table 2).

Since rate is distributed on a circle (with rates of 23 adjacent to rates of 0), the data violate the distributional assumptions of a linear model. For example, the circular von Mises distribution fit a mean rate of 4.7, while the intercept of the linear model $\beta_0$ was 5.75, reflecting a drift towards 11.5 at the middle of the $\{1, \ldots, 24\}$ interval. We tested the robustness of the model to this violation by fitting four additional models whose rates had been shifted uniformly to three different points on the interval,

$$(Rate_{i,subj,group} - 6) \bmod 24 = \beta_0 + \beta_{round} + \beta_{groupsize} + \beta_{condition} + (u_{subj} + u_{group} + e_{i,subj,group})$$
$$(Rate_{i,subj,group} + 6) \bmod 24 = \beta_0 + \beta_{round} + \beta_{groupsize} + \beta_{condition} + (u_{subj} + u_{group} + e_{i,subj,group})$$
$$(Rate_{i,subj,group} + 12) \bmod 24 = \beta_0 + \beta_{round} + \beta_{groupsize} + \beta_{condition} + (u_{subj} + u_{group} + e_{i,subj,group})$$
$$(Rate_{i,subj,group} + 18) \bmod 24 = \beta_0 + \beta_{round} + \beta_{groupsize} + \beta_{condition} + (u_{subj} + u_{group} + e_{i,subj,group})$$

Obviously, these models fit different values of the mean rate $\beta_0$. In all five models, the effect of round was significant, and the effects of group size and condition were insignificant (Table 3). The +0, +6, and +12 models fit comparable positive values to the coefficient $\beta_{round}$, all near 0.01. By contrast, the equivalent -6 and +18 models fit (the same) negative value to $\beta_{round}$. This is an understandable artifact; the distribution of rates will more flagrantly violate normality as its peak becomes centered at the "wraparound" point at 0 = 24 modulo 24. However, the distribution in the basic model (Rate + 0) should be robust to this violation of normality because its peak was far from the edges of the interval and the coefficient on even the strongest effect, $\beta_{round}$, was not large enough for rates to circumlocute their range. Over 200 rounds, $\beta_{round} = 0.0085$ corresponds to a total acceleration of 1.7 choices.

It is evident in visualizations of both the time and frequency domains that rotations accelerate over time (Figure 5), and the statistics support this conclusion. Analyzed within-subject and within-experiment, mean rate increased significantly over the 200 rounds of play ($\chi^2_1$=192, p<0.001; Table 2) by 1.7 choices. Group size and condition were not predictors of rate.

**Discussion**

The iterated elimination of non-rationalizable strategies behind Nash equilibrium is intended to mimic the human process of reasoning iteratively through the incentives of other people. By implication, increasing depths of human iterated reasoning are presumed to produce behavior increasingly consistent with Nash equilibria. Behavior in the Mod Game suggests that iterated reasoning may sustain periodic a behavior that does not converge to a fixed-point.

The entropy, earnings, and clustering of behavior in the Mod Game are inconsistent with the uniformly random play prescribed by the mixed Nash equilibrium, and other popular solution concepts (Camerer, Ho, & Chong, 2002b; McKelvey & Palfrey, 1995). Furthermore, the persistent periodicity in observed choices is inconsistent with any fixed-point solution concept. These findings appear to place rather severe constraints on possible explanations of behavior in the Mod Game. They are, however, consistent with an explanation suggested by the participants themselves. Although introspective reports must be interpreted with caution, participants described an iterated reasoning process driving rotations through belief space (Shepard & Metzler, 1971).

Going back to Selten, research on iterated reasoning works towards proving that that use of iterated reasoning implies greater fidelity with equilibrium predictions. This cannot be the case if iterated reasoning in the Mod Game is driving periodic behavior. In fact, if trajectories in belief space describe circles around the Nash equilibrium, the prescriptions of iterated reasoning are literally orthogonal to it, and iterated reasoning is complicit in the convergence of sophisticated reasoners towards a periodic attractor.

The heuristic learning direction theory is particularly promising for describing the individual reasoning process behind periodicity and group-level clustering. By this theory, participants learn to iterate through a limited $k$ number of steps of reasoning through the reasoning of others (Nagel, 1995; Selten, 1998). As they gain experience, participants make minor myopic adjustments to their current $k$—up or down depending on the direction of their error in the previous round. With time, participants learn the mean sophistication of their group, and they either adjust their own level of sophistication or heuristically adjust their rate of rotation to mimic a given level of sophistication. Because participants had only 6 seconds to decide, it isn't likely that they literally worked through the costly iteration process every round. But it also isn't necessary—as long participants thought others were using iterated reasoning, or thought others might think others were, a participant could use the visual layout of choices as a heuristic proxy to mimic the full iterated reasoning process. Iterated reasoning, even if it isn't the actual process driving decisions in the Mod Game, still gives the most compelling conceptual scheme for explaining how participants reasoned through it.

In the Mod Game, participants preferred rates of 1–11 to rates of 13–23 (by a 3:1 ratio). Why is there such a strong regularity in rates across experimental sessions? There would be no such limit if participants used a theory-free empirical time-series method to learn their group's emergent rate. But rates grounded in iterated reasoning would be expected to show precisely the limits observed. Camerer and Ho fit over one hundred games to an iterated reasoning model and found that a degree of ~1.5 thinking steps fit the best, and that most games elicit a range of 0–3 steps (Camerer, Ho, & Chong, 2004). From the perspective of iterated reasoning, the observed mean rate of 4.7 choices per round corresponds to 2.35 thinking steps—a hop of two choices for each additional level $k$—within the range of $k$'s observed in other experiments. A problem with applying iterated reasoning to an intransitive game is that we must assume that a 0-step reasoner preserves the previous round's choice. This assumption is difficult to defend without resorting to more exotic behaviors, like the default heuristic (by which participants repeat their previous action when they lack a reason to change it) or strategic teaching (by which sophisticated participants "play dumb" to manipulate unsophisticated players into some favorable pattern of coordination) (Camerer, Ho, & Chong, 2002a; Gigerenzer & Todd, 1999). But research on thinking-steps can also account for the mean acceleration of 1.7 choices per round over 200 rounds. 1.7 choices would correspond to 0.85 thinking steps, well within the increase of 0.5–1 thinking-step increase observed in other experiments (Camerer, 2003, p. 322; Duffy & Nagel, 1997; Stahl, 1999).

Iterated reasoning is an active research subject, but researchers downplay the importance of the heuristic adjustment process that originally accompanied it (Costa-Gomes, Crawford, & Iriberri, 2009; Crawford & Iriberri, 2007b; T. H. Ho & Su, 2011; Kawagoe & Takizawa, 2008; Stahl, 1999). However, the adjustments of learning direction theory are necessary to explain why 49% of non-zero adjustments to rate in the Mod Game were decelerations. Learning direction theory also provides an individual-level mechanism for group-level clustering (S. Frey & Goldstone, 2011).

Dynamical systems and statistical mechanics offer powerful tools for characterizing the types of complex emergent patterns that we observe here. Intransitive dominance relations between distributed mobile agents have been shown to foster periodic dynamics universally (Reichenbach et al., 2007; Reichenbach, Mobilia, & Frey, 2008; Winkler, Reichenbach, & Frey, 2010). And in the Mod Game, clustering and periodicity may both fall out of a dynamic analogous to that driving the synchronization of systems of coupled oscillators (Mirollo & Strogatz, 1990). Specifically, clustering and convergence on a mean rate can be treated as phase-locking and frequency-locking, respectively.

Generally, a satisfactory model of behavior in the Mod Game will make adjustments around a time-dependent rate, inducing non-stationary dynamics through a regime of stable cyclic attractors that captures both the persistent periodicity and the changes in rate over the course of the experiment. Limit cycles are the non-fixed-point attractors that have received the most attention in game theory, and the observed periodic behavior is qualitatively consistent with this type of dynamic. But periodicity is also consistent with other dynamics, like quasi-cycles, quasi-periodic oscillations, some chaotic attractors, and even very slow cyclic transients towards a fixed point (Kantz & Schreiber, 2003; Turchin & Taylor, 1992).

Most groups that played the Mod Game can be described as clustering and cycling stably at a slowly increasing rate. Qualitatively, there were some exceptions to the general trend. The middle columns of Figures S2-S9 show rates over time for 29 sessions. Most groups exhibit coordination on a rate between 1 and 12 after some transient. Group 3 showed a particularly long transient. Groups 7 and 9 exhibit rates that are difficult to distinguish from random. Clustering in discarded group 3 seems to dissolve half way through the experiment. Participants in discarded groups 5 and 7 seemed to converge to pure strategies. The most interesting exceptions were groups 10 and 12, which exhibited persistent clustering and cycling, but at much higher rates than those observed in any other group. Group 12 settled at a rate of 12, and Group 10 continued accelerating through the entire range of rates, such that they were rotating in the "wrong" direction by the end of the experiment. Overall, we do not make a strong claim as to whether rates stabilize or increase indefinitely. There seems to be heterogeneity between groups, with some converging upon a stable rate of rotation, and others continuing to accelerate through the whole session.

The description of the subject pool mentions eight participants that made large numbers of repeated choices or null choices and that were excluded from the analysis. Including these participants does not affect the main results of this manuscript: rate of approximately 4, increasing significantly, and driving periodicity that registers a significant spike in a Fourier analysis. The biggest effect of including all of the data is in the polar histogram of rates—the second panel of Figure 3—which registers a larger spike at rate zero. But were participants sufficiently motivated? While participants were paid, rates were below the standard for economics laboratories; expected earnings were 1¢ per point and mean earnings were $1.33 over ~30 minutes. Undermotivated behavior is traditionally invoked to explain deviations from predicted fixed-point behavior. Still, a constructive theory would be necessary to explain why these deviations were not to a poorly defined fixed point, but to a more profitable higher-dimensional attractor that has been anticipated for 50 years. Within some dynamical frameworks, limit cycle regimes are more prominent in games with lower payoffs (Galla & Farmer, 2011; Sato & Crutchfield, 2003). In this context, motivation is not a mere methodological nuisance (Camerer & Hogarth, 1999; Smith & Walker, 1993), but a theoretically grounded concept whose manipulations make substantive predictions, predictions that our work supports.

Accepting the coexistence of iterated reasoning and periodic behavior does not fix all of the problems presented by this work. Existing models of complex learning dynamics cannot account for important features of periodicity in the Mod Game. If participants' beliefs are traversing a limit cycle regime, these cycles are different from any that have been predicted. Participants choose their next move using a conception of rate that leads them to "hop" around the circle of choices. As groups, they coordinate their hopping and cluster around specific choices. Neither of these behaviors has been predicted in the dynamics of game learning. Additionally, participants' rates increase significantly over time, reflecting either convergence, in a non-stationary stochastic system, to a periodic attractor that is changing shape, or the ephemeral behavior of trajectories that are converging only slowly to a stationary periodic attractor.

**Conclusions**

We have used the Mod Game, an *n*-person generalization of Rock-Paper-Scissors, to document the emergence of a stable, profitable periodicity in group behavior. We argue that the interactions between bounded individuals led groups to cluster and cycle through the space of choices. These cycles reflect periodic trajectories through the space of participants' probability vectors. In people, these trajectories can only be inferred via observable behavior, so we cannot offer more direct support for the hypothesis that participants' beliefs have converged upon a regime of stable periodic trajectories.

Cycles in the belief space of learning agents have been predicted for many years, particularly in games with intransitive dominance relations, like Matching Pennies and Rock-Paper-Scissors, but experimentalists have only recently started looking to these dynamics for experimental predictions. This work should function to caution experimentalists of the dangers of treating dynamics as ephemeral deviations from a static solution concept. Periodic behavior in the Mod Game, which is stable and efficient, challenges the preconception that coordination mechanisms must converge on equilibria or other fixed-point solution concepts to be promising for social applications. This behavior also reveals that iterated reasoning and stable high-dimensional dynamics can coexist, challenging recent models whose implementation of sophisticated reasoning implies convergence to a fixed point (Camerer, Ho, & Chong, 2002a). Applied to real complex social systems, this work gives credence to recent predictions of chaos in financial market game dynamics (Galla & Farmer, 2011). Applied to game learning, our support for cyclic regimes vindicates the general presence of complex attractors, and should help motivate their adoption into the game theorist's canon of solution concepts.

**Acknowledgements**

This research was supported in part by National Science Foundation IGERT training grant 0903495 in the Dynamics of Brain-Body-Environment Systems at Indiana University, National Science Foundation REESE grant 0910218, and Department of Education IES grant R305A1100060. Contact sethfrey@indiana.edu for the raw data. The authors would like to thank Benjamin Marchus for his contributions to this project, as well as Ryan Murphy, Peter Todd, Yuzuru Sato, Tom Wisdom, Jerome Busemeyer, and Jay Silver for insight into the analysis and interpretation.

**Figure Legends**

**Figure 1**

**Experiment interface.** This screenshot was taken during a pilot *increment* session, after all decisions had been submitted, and as all decisions and rewards in a round were being reported. Participants saw their own choices as the red 'X'. Previous experiments have tested the same rule with visual arrangements besides the circle (S. Frey & Goldstone, 2011). See Video S1 for the complete video for a typical session.

**Figure 2**

**Observed mean entropy, efficiency, and distance compared to random.** The boxes report means of observed behavior with bootstrapped 99% confidence intervals. The crosses give values expected from uniformly random behavior.

**Figure 3**

**Distributions of observed choices, rates, and accelerations.** The **top panel** compares distributions over the twenty-four choices, over increment and decrement conditions, against a random baseline. Without temporal information, aggregated choices are difficult to distinguish from uniformly random behavior. The **middle panel** compares distributions of participant rates. The observed distribution is consistent with the measured mean rate of 4.7 choices per round, forward or backward for increment and decrement conditions, respectively. The **bottom panel** illustrates accelerations (the difference between consecutive first differences). Observed accelerations are consistent with behavior that either maintains the previous round's rate or makes only minor adjustments to it. Note that, since the null hypothesis is identical across measures, the circles representing random behavior in each panel have identical radius.

**Figure 4**

**Aggregated frequency spectra of participant time series, with baseline and predictions.** The frequency spectra for the first and second 100 rounds of the experiment show the development of cycles. For consistency, the horizontal axis is in units of rate rather than frequency. The frequency spectrum shows a prominent spike in the latter half of the experiment, corresponding to a rate of rotation of about 7 choices per round. This spectrum is the aggregate of spectra from many statistically independent sessions. To control for artifacts and maintain independence, the data were transformed and resampled before transformation to the frequency domain. The dark vertical bar illustrates the spike location predicted by the mean rate. The lighter bars give predictions for mean rates calculated using only the first (left) and second (right) 100 rounds of play.

**Figure 5**

**Mean rates in time and aggregate spectrogram.** The **left** panel shows the mean rate in each group, at each round. The **right** panel shows a spectrogram (with a window size of 20 rounds) for resampled observed data. These figures show changes in rotation over the sequence of 200 rounds of play. In the spectrogram, the brightness of a pixel indicates the amplitude of the corresponding frequency component. These panels show statistically significant increases in the rate of periodic behavior, in both the time and frequency domains.

**Supporting Figure Legends**

**Figure S1**

**Video of experimental session.** This video shows a screen capture of session #21, sped up approximately 30 times.   While some groups never exhibited periodic behavior, this group was typical of that majority that did. Though this video is taken from the experimenter's perspective, participants saw nearly the same view, including complete information after every round as to every player's position and earnings.   The differences are that they had feedback on their accumulated earnings, and they saw their own icon as a red 'X'.

**Figure S2**

**Choice, rate, and acceleration data for groups 1–4.** Within each row of three graphs, dot colors designate the same group member.

**Figure S3**

**Choice, rate, and acceleration data for groups 5–8.**

**Figure S4**

**Choice, rate, and acceleration data for groups 9–12.**

**Figure S5**

**Choice, rate, and acceleration data for groups 13–16.**

**Figure S6**

**Choice, rate, and acceleration data for groups 17–20.**

**Figure S7**

**Choice, rate, and acceleration data for groups 21–22.**

**Figure S8**

**Choice, rate, and acceleration data for discarded groups 1–4**

**Figure S9**

**Choice, rate, and acceleration data for discarded groups 5–7**

**Tables**

**Table 1: Summary of experimental sessions**

| Session number | Group size | Condition | Time, Date |
|---|---|---|---|
| 1 | 7 | decrement | 11:00, 2011/09/15 |
| 2 | 3 | decrement | 12:00, 2011/09/30 |
| 3 | 10 | increment | 13:00, 2011/09/15 |
| 4 | 5 | increment | 11:00, 2011/09/22 |
| 5 | 2 | decrement | 11:00, 2011/09/23 |
| 6 | 6 | decrement | 11:00, 2011/09/16 |
| 7 | 9 | increment | 12:00, 2011/09/08 |
| 8 | 8 | decrement | 12:00, 2011/09/09 |
| 9 | 3 | increment | 12:00, 2011/09/14 |
| 10 | 9 | decrement | 11:00, 2011/09/09 |
| 11 | 8 | decrement | 13:00, 2011/09/08 |
| 12 | 7 | increment | 12:00, 2011/09/16 |
| 13 | 2 | decrement | 10:00, 2011/09/30 |
| 14 | 3 | increment | 11:00, 2011/09/21 |
| 15 | 8 | decrement | 11:00, 2011/09/08 |
| 16 | 2 | increment | 11:00, 2011/10/07 |
| 17 | 5 | increment | 11:00, 2011/12/05 |
| 18 | 6 | decrement | 09:00, 2011/12/05 |
| 19 | 8 | increment | 12:00, 2011/12/01 |
| 20 | 3 | increment | 11:00, 2011/11/30 |
| 21 | 3 | decrement | 12:00, 2011/11/17 |
| 22 | 5 | decrement | 12:00, 2011/11/10 |
| Discard 1 | 8 | increment | 12:00, 2011/09/15 |
| Discard 2 | 9 | increment | 15:00, 2011/12/07 |
| Discard 3 | 9 | decrement | 14:00, 2011/12/07 |
| Discard 4 | 7 | increment | 10:00, 2011/12/05 |
| Discard 5 | 5 | increment | 11:00, 2011/12/01 |
| Discard 6 | 4 | increment | 11:00, 2011/11/18 |
| Discard 7 | 3 | decrement | 11:00, 2011/11/16 |

**Table 2: Linear effects on rate, with random effects for subject and session.**

|  | coefficient | df | LL | $\chi^2$ | $\chi^2$ df | *p value* |
|---|---|---|---|---|---|---|
| full model |  | 7 | -76579 |  |  |  |
| intercept | **5.75** | 6 | -76591 | 23.9 | 1 | <0.001 |
| round | **0.00854** | 6 | -76675 | 192 | 1 | <0.001 |
| group size | 0.239 | 6 | -76581 | 2.51 | 1 | 0.113 |
| condition | 0.269 | 6 | -76580 | 0.41 | 1 | 0.522 |

This table reports effects of $\chi^2$ tests on reduced models. **Bold** coefficients are significant.

**Table 3: Robustness of the linear model given nonlinear (circular) rate.**

|  | $\beta_0$ | $\beta_{round}$ | $\beta_{groupsize}$ | $\beta_{condition}$ |
|---|---|---|---|---|
| *Rate +0 (base model)* | **5.75** | **0.00854** | 0.239 | 0.269 |
| *Rate +6* | **7.25** | **0.0156** | 0.422 | 0.239 |
| *Rate +12* | **12.5** | **0.00776** | -0.196 | -0.266 |
| *Rate +18, Rate − 6* | **20.6** | **-0.0136** | -0.925 | -0.268 |

**Bold** coefficients are significant at $p < 0.001$. Other coefficients are not significant below $p < 0.05$. With rate defined on a lattice of diameter 24, Rate + 18 = Rate − 6.

# Figures

## Figure 1

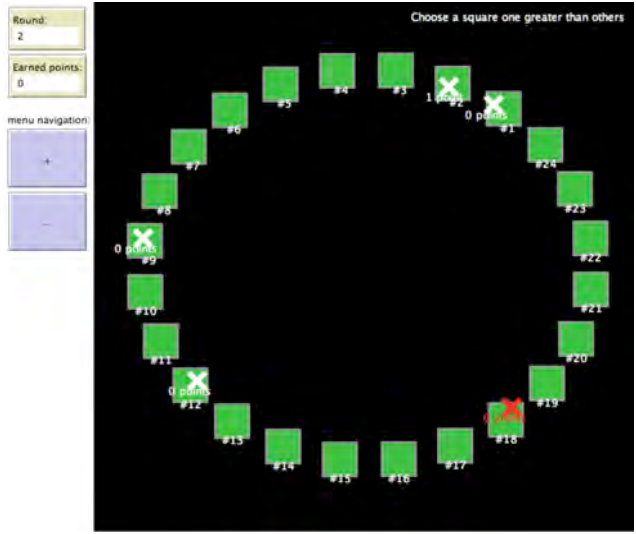

**Figure 2**

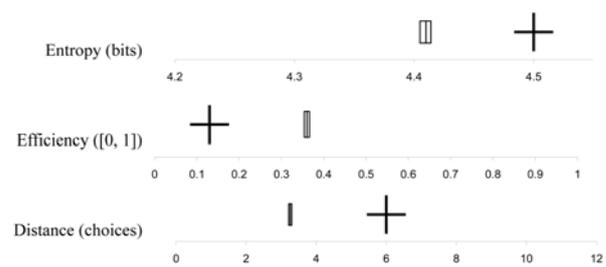

**Figure 3**

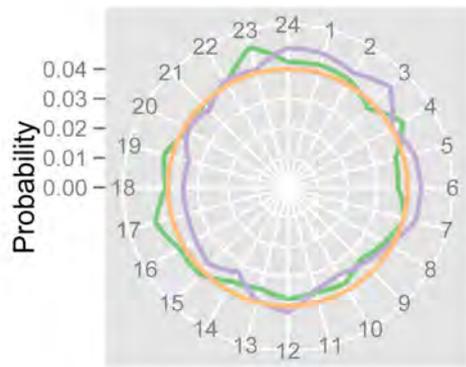

Participant choices

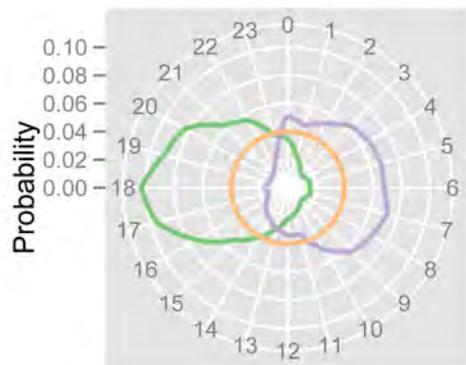

Participant rates

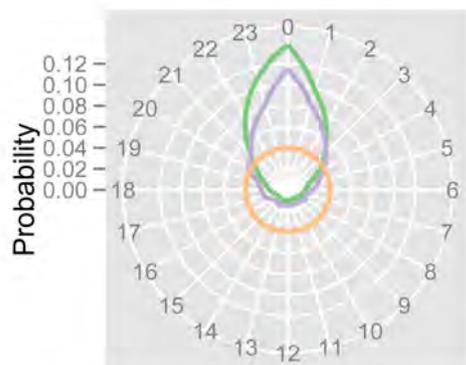

Participant accelerations

**Participant**
— Decrement: Observed
— Incremental: Observed
— Random

**Figure 4**

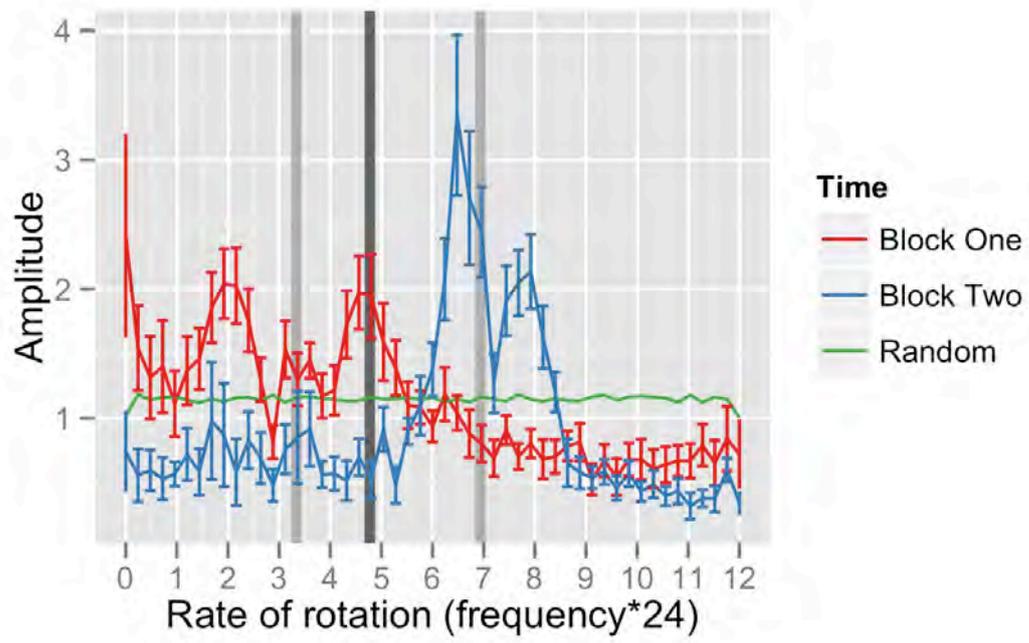

**Figure 5**

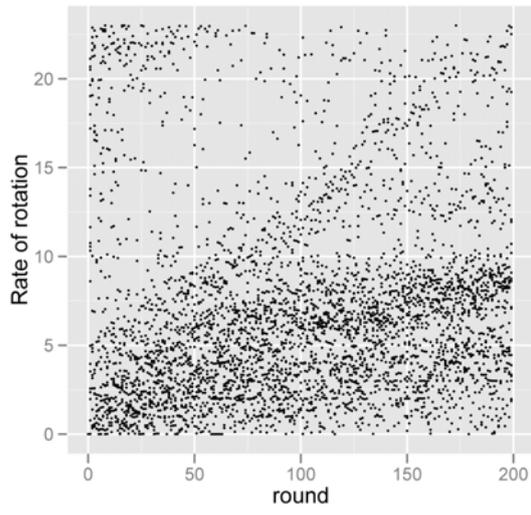 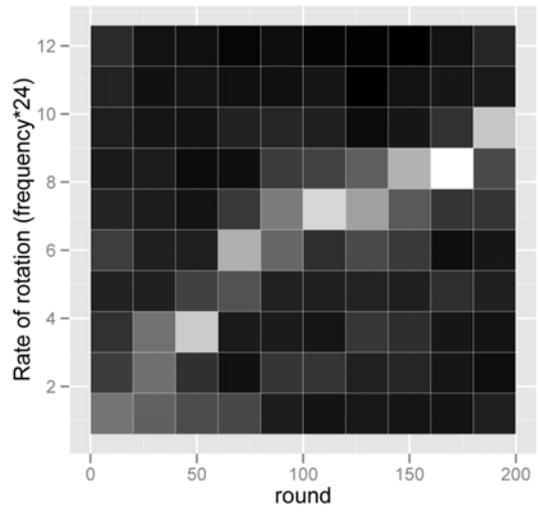

**Supporting Figures**

**Figure S1**

https://vimeo.com/50459678

**Figure S2**

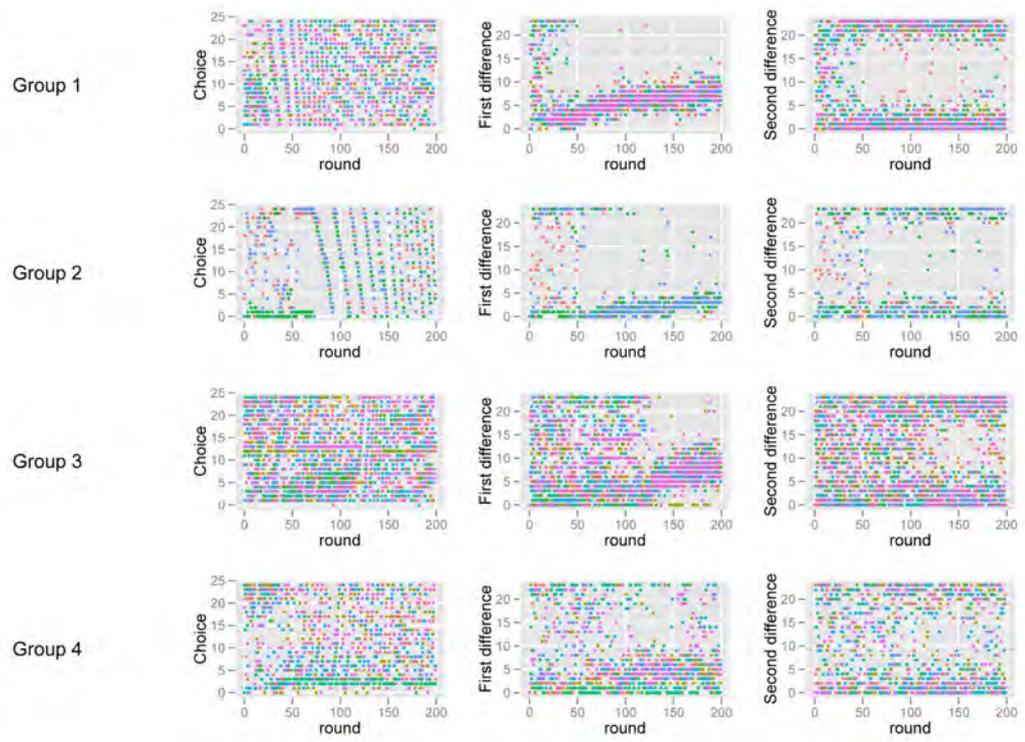

**Figure S3**

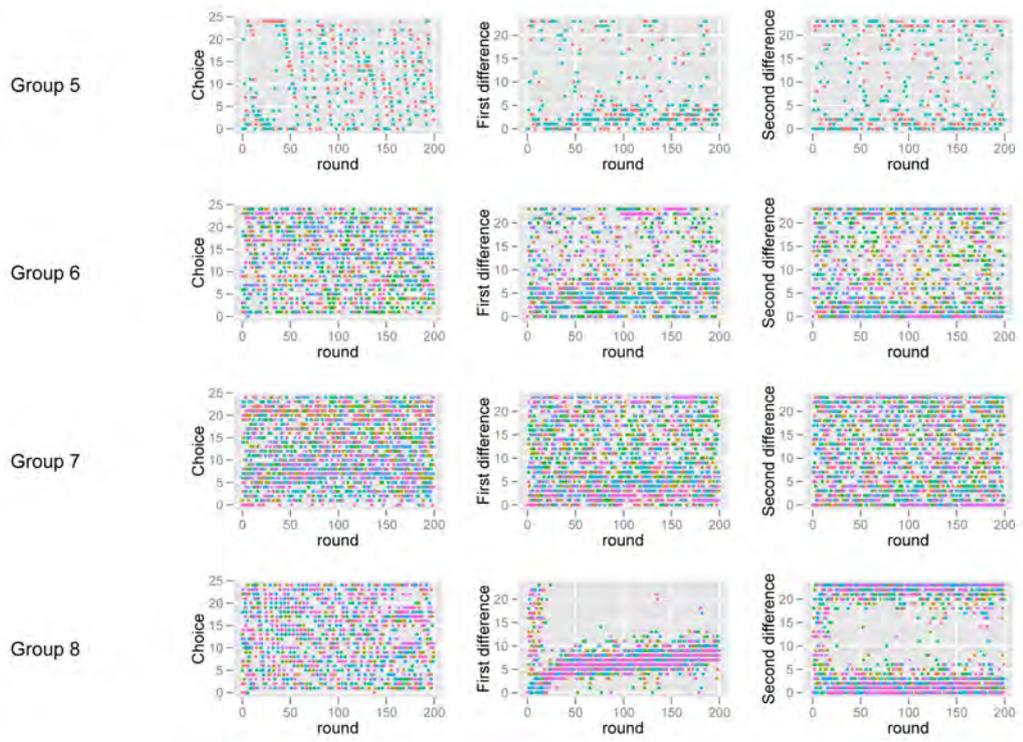

**Figure S4**

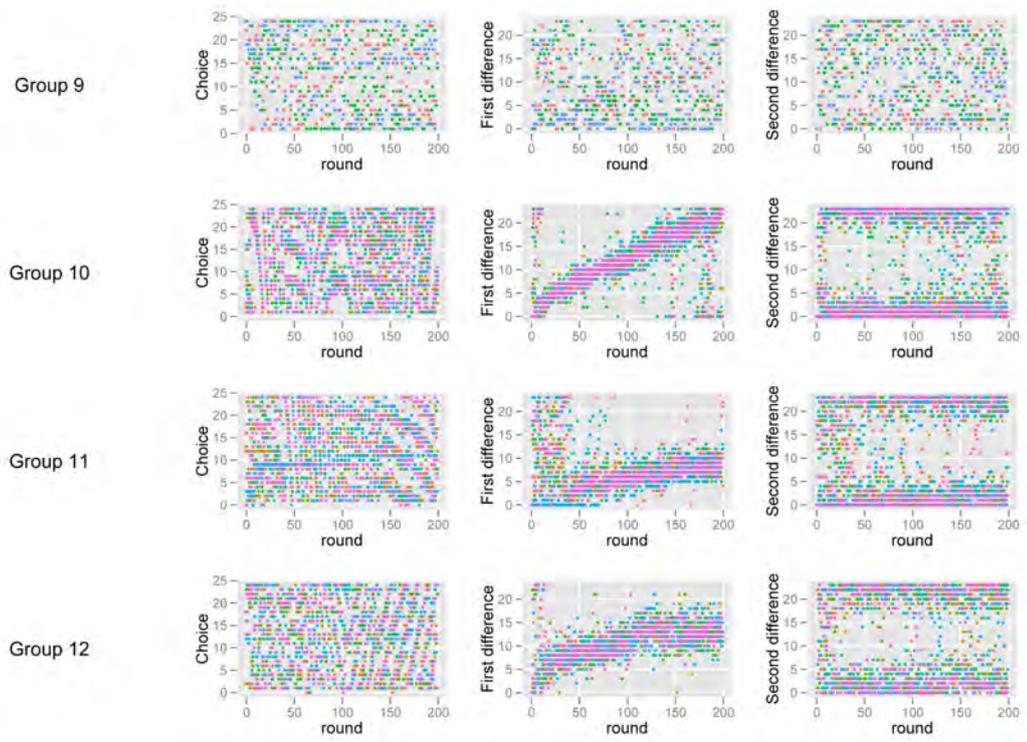

**Figure S5**

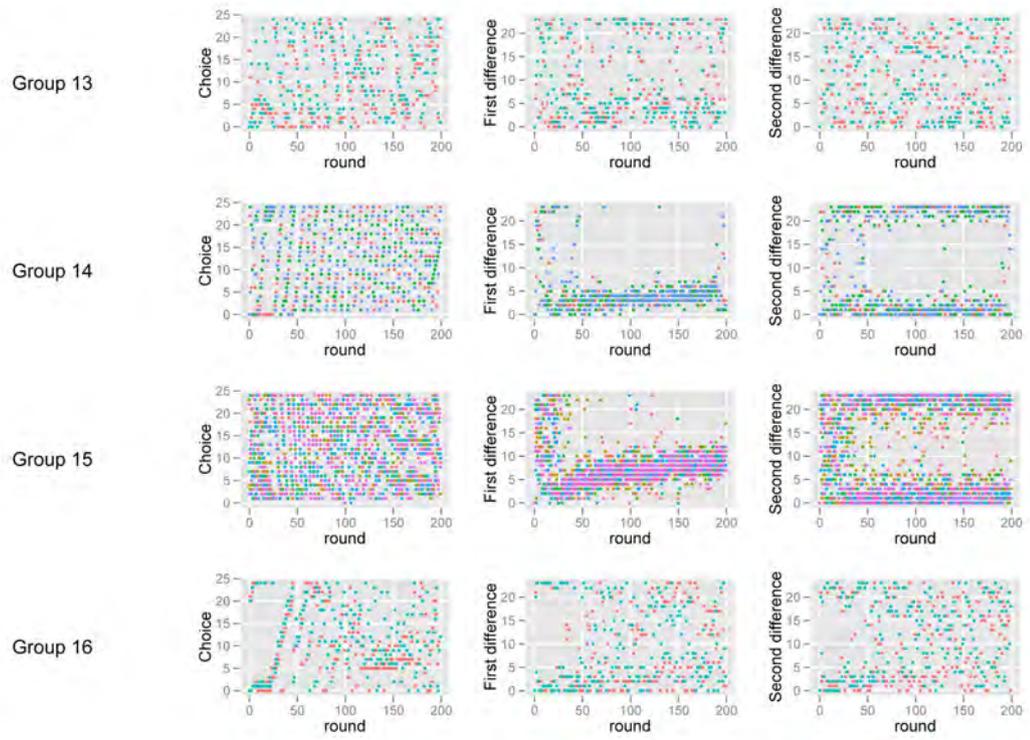

**Figure S6**

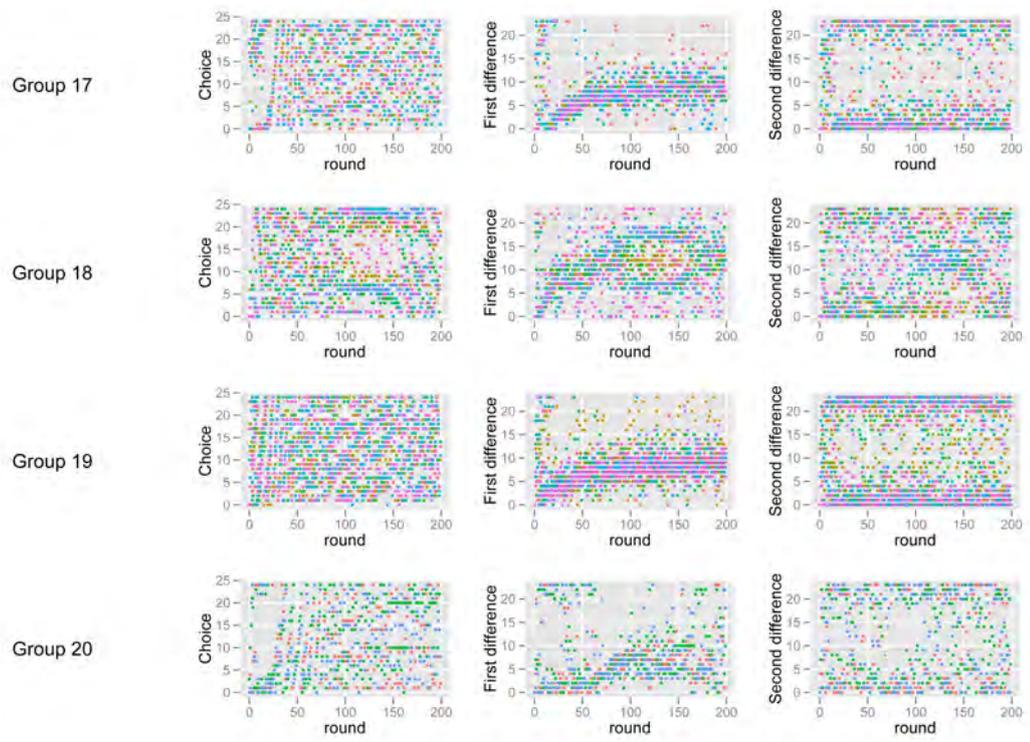

**Figure S7**

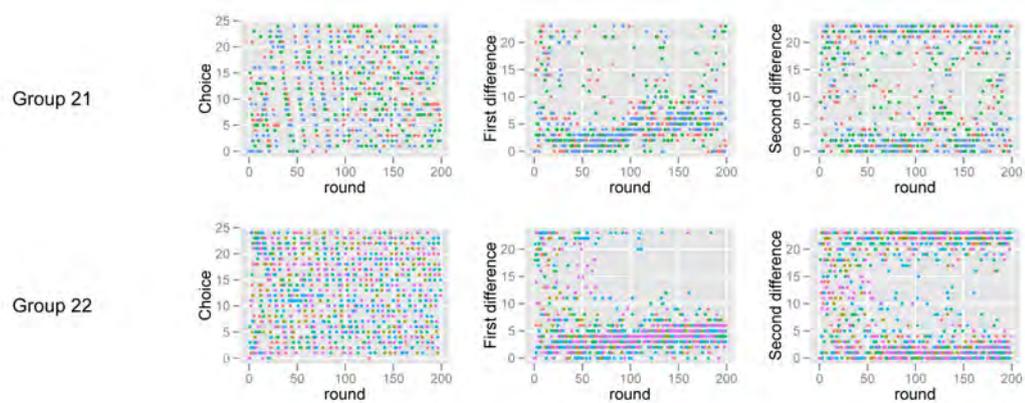

**Figure S8**

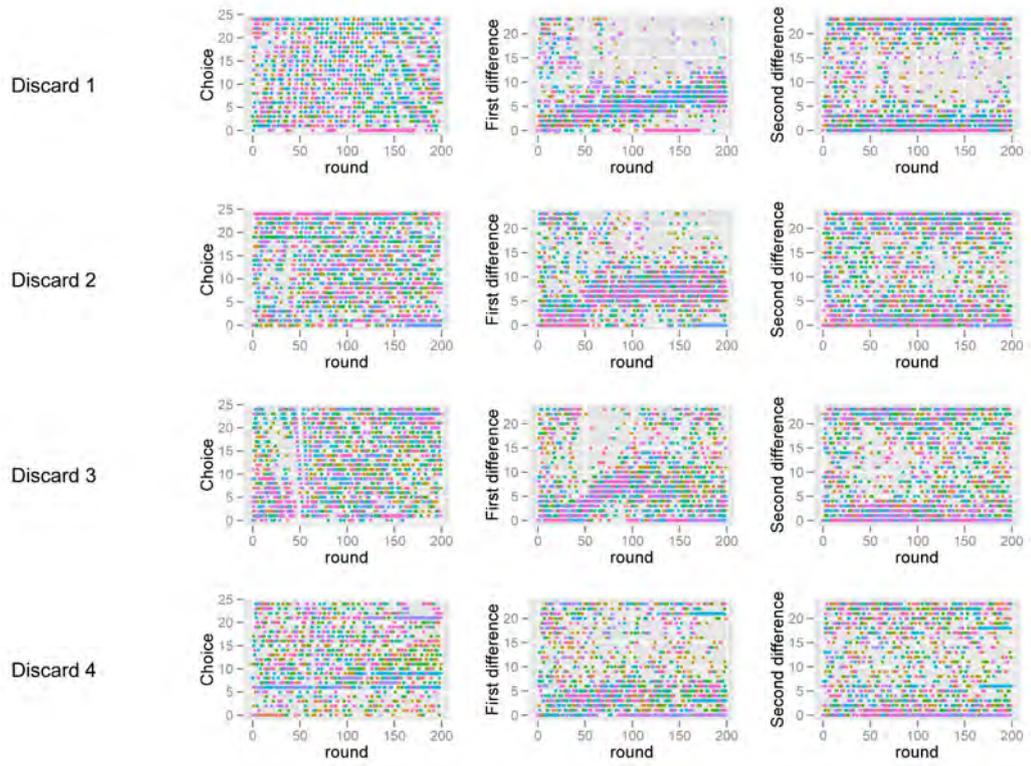

**Figure S9**

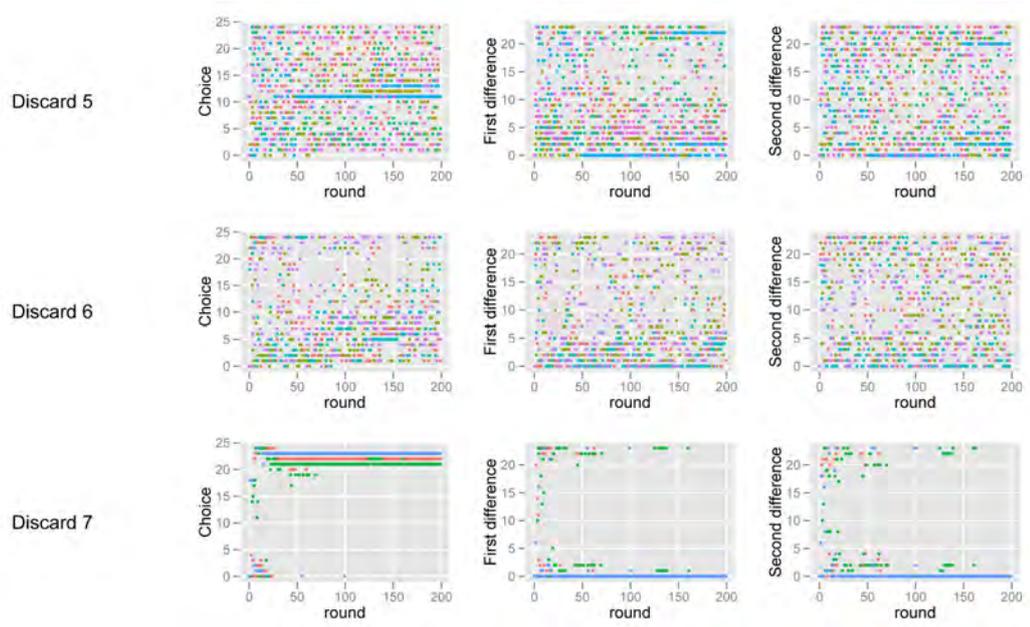